\documentclass[submit]{eps}
\usepackage{times}
\usepackage{graphicx}
\newcommand       \Angstrom     {\,{\rm \AA}}

\newcommand       \g          {\,{\rm g}}
\newcommand       \cm          {\,{\rm cm}}
\newcommand       \um           {\,{\rm \mu m}}
\newcommand       \mum           {\,{\rm \mu m}}
\newcommand       \Rv           {{R_{\rm V}}}
\newcommand       \Av           {{A_{\rm V}}}

\newcommand       \Ks           {{\rm K_{s}}}

\newcommand       \simali       {{\sim}\,}
\newcommand       \magni        {\,{\rm mag}}
\newcommand       \ab           {a_{\rm b}}
\newcommand       \nH           {n_{\rm H}}
\newcommand       \NH           {N_{\rm H}}

\newcommand       \Asil           {A_{\rm si}}
\newcommand       \alphasil     {\alpha_{\rm si}}
\newcommand       \absil          {a_{\rm b,si}}
\newcommand       \Acarb         {A_{\rm c}}
\newcommand       \alphacarb   {\alpha_{\rm c}}
\newcommand       \abcarb        {a_{\rm b,c}}
\newcommand       \Vsil             {V_{\rm tot,si}}
\newcommand       \Vcarb          {V_{\rm tot,c}}
\volumenumber{58}
\publishyear{2013}
\frompage{\pageref{firstpage}}
\topage{\pageref{finalpage}}

\shorttitle{Gao, Li, \& Jiang: Modeling the Extinction towards the GC}

\title{Modeling the Infrared Extinction toward the Galactic Center}
\author{Jian Gao$^1$, Aigen Li$^2$, and B.~W. Jiang$^1$}
\affiliation{$^1$Department of Astronomy, Beijing Normal University,
                 Beijing 100875, China;\\
             $^2$Department of Physics and Astronomy,
                  University of Missouri, Columbia, MO 65211, USA;\\
             }
\received{xxxx xx, 2012}\revised{xxxx xx, 2013}\accepted{May 29, 2013}
\abstract{
We model the $\simali$1--19$\mum$ infrared (IR) extinction curve
toward the Galactic Center (GC) in terms of
the standard silicate-graphite interstellar dust model.
The grains are taken to have a power law size distribution
with an exponential decay above some size.
The best-fit model for the GC IR extinction constrains
the visual extinction to be $\Av$\,$\simali$38--42$\magni$.
The limitation of the model, i.e., its difficulty in
simultaneously reproducing
both the {\sl steep} $\simali$1--3$\mum$ near-IR extinction
and the {\sl flat} $\simali$3--8$\mum$ mid-IR extinction
is discussed.
We argue that this difficulty could be alleviated by attributing
the extinction toward the GC to a combination of dust in different
environments: dust in diffuse regions (characterized by small $\Rv$
and steep near-IR extinction), and dust in dense regions (characterized
by large $\Rv$ and flat UV extinction).
}

\begin{document}
\label{firstpage}
\maketitle
\copyrighttext{}

\section{Introduction}
The wavelength dependence of the interstellar extinction
-- known as the ``interstellar extinction law (or curve)''
-- is one of the primary sources of information
about the interstellar grain population (Draine 2003).
The Galactic interstellar extinction curves in the ultraviolet (UV)
and visual wavelengths vary from one sightline to another,
and can be parameterized in terms of the single parameter
$\Rv \equiv \Av/E(B-V)$,
the total-to-selective extinction ratio (Cardelli et al.\ 1989).\footnote{%
   $^{1}$ $E(B-V)\equiv A_B - A_V$ is the interstellar reddening,
   $A_B$ is the extinction at the ``$B$'' (blue;
   $\lambda_B\approx4400\Angstrom$) band,
   and $A_V$ is the extinction at the ``$V$'' (visual;
   $\lambda_V\approx5500\Angstrom$) band.
   }
Larger values of $\Rv$ correspond to size distributions
skewed toward larger grains (e.g., dense clouds tend to have
large values of $\Rv>4$).
On average, the dust in the diffuse interstellar medium (ISM)
corresponds to $\Rv\approx3.1$.

However, the infrared (IR) interstellar extinction law,
which also varies from sightline to sightline, cannot be
simply represented by $\Rv$.
Various recent studies have shown that there does not
exist a ``universal'' near-IR (NIR) extinction law
(Fitzpatrick \& Massa 2009; Gao et al.\ 2009; Zasowski et al.\ 2009) and
the mid-IR (MIR) extinction law shows a flat curve
and lacks the model-predicted pronounced minimum extinction
around 7$\um$ (Draine 1989).\footnote{%
    $^{2}$ In this work by ``NIR'' we mean $1\mum < \lambda < 3\mum$
    and by ``MIR'' we mean $3\mum < \lambda < 8\mum$.
    }
It is worth noting that the flat MIR extinction curves
determined for various sightlines all appear to agree with
the extinction predicted by the standard silicate-graphite
interstellar grain model for $\Rv$\,=\,5.5
(Weingartner \& Draine 2001) (hereafter WD01),
which indicates a dust size distribution favoring larger sizes
compared to that for $\Rv=3.1$.

Recently, using the hydrogen emission lines of the minispiral
observed by ISO-SWS and SINFONI,
Fritz et al. (2011) derived the IR extinction curve
toward the inner GC from 1 to 19$\um$.
The extinction curve shows a {\sl steep} NIR extinction
consistent with that of Nishiyama et al. (2006, 2009)
and a {\sl flat} MIR extinction
consistent with other sightlines
(see Figure~1).
It differs from the IR extinction law toward the GC
derived by Rieke \& Lebofsky (1985) (hereafter RL85) and Rieke et al.\ (1989).
Based on their observations, Fritz et al.\ (2011)
argued that the extinction at the visual band ($\Av$) toward
the GC may be as high as $\Av$\,$\simali$59$\magni$
(with the exact $\Av$ depending on the chosen gas-to-dust ratio
$N_{\rm H}/\Av$), much larger than $\Av\sim31$
estimated by Rieke et al.\ (1989) which is commonly adopted in
the astronomical literature.

In this work, we try to use the standard interstellar grain model
which consists of graphite and silicate grains (Draine \& Lee 1984)
to fit the observed IR extinction curve toward the GC of
Fritz et al. (2011) and constrain the total optical extinction ($\Av$)
toward the GC. \S2 briefly describes the grain model.
Our model results are presented in \S3 and discussed in \S4.
In \S5 we summarize the major conclusion of this work.

\begin{figure}[t]
\centerline{\includegraphics[angle=0,scale=0.5,clip]{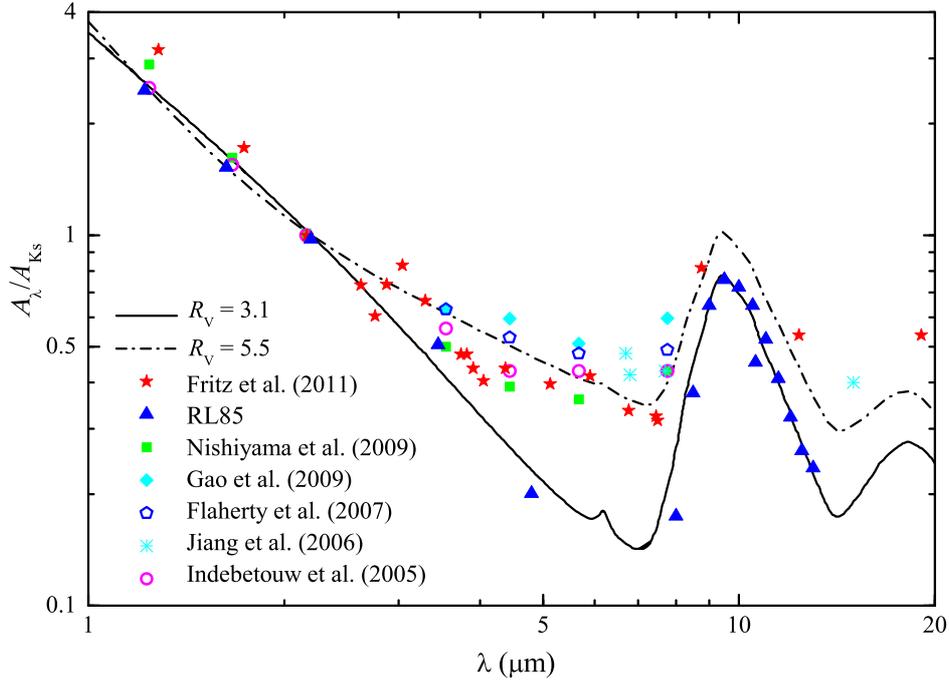}}
\caption{
         IR extinction laws compiled from the literature.
         Red stars plot $A_{\lambda}/A_\Ks$
                   toward the GC based on H lines (Fritz et al. 2011).
         Blue triangles are derived from stars
                   toward the GC (RL85).
         Green squares are derived from the red clump giants
                   toward the GC (Nishiyama et al.\ 2009).
         Cyan diamonds are the Galactic plane average extinction
                   at $|l| < 5^{\circ}$ and  $|b| < 2^{\circ}$ (Gao et al.\ 2009).
         The other three kinds of symbols plot the extinction laws
                   obtained from sightlines away from the GC.
         For comparison, the extinction curves calculated
                   from the interstellar grain model (WD01)
                   for $\Rv = 3.1$ (black solid line) and
                   $\Rv = 5.5$ (black dot-dashed line)
                   are also shown.
         \label{fig1}}
\end{figure}

\section{Dust Model}\label{model}
%

We take the dust to be a mixture of separate amorphous silicate
and graphite grains, with the optical properties taken from
Draine \& Lee (1984). For the dust size distribution, we adopt
a power law with an expoential cutoff at some large size:
$dn/da = A\,\nH\,a^{-\alpha} \exp\left(-a/\ab\right)$
with $50\Angstrom < a < 1\um$,
where $a$ is the grain radius,\footnote{%
  $^{3}$ We assume the dust to be spherical.
  }
$dn$ is the number density of dust with radii
in the interval [$a$, $a$\,$+$\,$da$] per H nuclei,
$\nH$ is the number density of H nuclei,
$A$ is the normalization constant,
$\alpha$ is the power index, and $\ab$ is the cutoff size.
In our modeling, we will have six parameters:
$\Asil$, $\alphasil$, $\absil$ for the silicate component,
and $\Acarb$, $\alphacarb$, $\abcarb$ for the graphite component.
The total extinction at wavelength $\lambda$ is given by
\begin{equation}\label{ext}
A_\lambda =(2.5\log e) \NH \sum_i \int da\frac{dn_i(a)}{da}
           C_{{\rm ext},i}(a,\lambda), \\
\end{equation}
where the summation is made over the two grain types
(i.e., silicate and graphite), $\NH\equiv\int\nH\,dl$
is the H column density which is the H number density
integrated over the line of sight $l$,
and $C_{{\rm ext},i}(a,\lambda)$
is the extinction cross section of grain type $i$ of size $a$
at wavelength $\lambda$.
The goodness of fitting is evaluated by 
\begin{equation}\label{error}
\frac{\chi^2}{\rm d.o.f} =
\frac{\sum_{j=1}^{N_{\rm obs}}
\left(A_{\lambda}^{\rm mod} - A_{\lambda}^{\rm obs}\right)^2/\sigma_j^2}
{N_{\rm obs}-N_{\rm para}} ~~,
\end{equation}
where $A^{\rm obs}_{\lambda}$ is the IR extinction toward the GC
derived by Fritz et al.\ (2011) (see their Table~2),
$N_{\rm obs}$ is the number of observational data points,
$N_{\rm para}$ is the number of adjustable parameters
($N_{\rm para} =6$ if we assume different size distributions
for silicate and graphite; $N_{\rm para} =4$ if we assume that
both dust components have the same size distribution),
$A^{\rm mod}_{\lambda}$ is the model extinction computed from
eq.1, and $\sigma_j$ is the weight of the observed extinction.
Assuming that $\approx$30\% is in the gas phase,
WD01 adopt the solar abundance of Grevesse \& Sauval (1998) to constrain their models.
Their CASE A models tried to seek the best fit by varying the total volume per H in both the
carbonaceous and silicate distributions, while their CASE B models fixed at
approximately the values found for $\Rv=3.1$.
Following WD01, we fix the total dust quantity (per H nuclei) to
be consistent with the cosmic abundance constraints.
Let $\Vsil$ be the total volume of the silicate dust,
and $\Vcarb$ be the total volume of the graphitic dust.
We take $\Vsil = 2.98\times10^{-27} \rm\,cm^{3}\,H^{-1}$
and $\Vcarb = 2.07 \times 10^{-27} \rm\,cm^{3}\,H^{-1}$
(i.e., values for constraining all ``CASE A'' models of WD01)\footnote{%
$^{4}$
The abundance of carbonaceous and silicate given by Asplund et al. (2009) are
$2.95\times10^{-4}$ and $3.55\times10^{-5}$, respectively. If considering the solar abundance of
Asplund et al. (2009), one would get $\Vsil = 2.91\times10^{-27} \rm\,cm^{3}\,H^{-1}$
and $\Vcarb = 1.85 \times 10^{-27} \rm\,cm^{3}\,H^{-1}$, i.e. $\Vsil/\Vcarb=0.61/0.39$, which is
close to the ratio of the WD01 ``CASE B'' models. We also did fit the extinction curve by varying
the ratio of $\Vsil/\Vcarb$: by taking the silicate-to-graphite mass ratio to $m_{\rm gra}/m_{\rm sil}=$ 0.4, 0.5, and 0.6,
our model results show that $\Av$ toward the GC is in the range of $\simali$35--45$\magni$.
     }.
We will also consider $\Vsil = 3.9\times10^{-27} \rm\,cm^{3}\,H^{-1}$
and $\Vcarb = 2.3 \times 10^{-27} \rm\,cm^{3}\,H^{-1}$
(i.e., fixed values for all ``CASE B'' models of WD01).\footnote{%
    $^{5}$ The WD01 ``CASE B'' model extinction curve of $\Rv =5.5$ shows a similar tendency as the observed flat MIR extinction
    (Draine 2003; Indebetouw et al.\ 2005; Jiang et al.\ 2006; Gao et al.\ 2009; Zasowski et al.\ 2009; Nishiyama et al.\ 2009).
     }
The mass densities of amorphous silicate and graphite are taken to be
$\rho_{\rm sil} \approx 3.5\g\cm^{-3}$ and $\rho_{\rm carb} \approx
2.24\g\cm^{-3}$.

\begin{table}[t] 
\renewcommand{\arraystretch}{1.2}
\vspace{-.3cm}
\caption{Model parameters for fitting the GC IR extinction curve}
\vspace{-.1cm}
\begin{center}
\begin{tabular}{cccccrcc}
\hline \hline
Abundance &  \multicolumn{2}{c}{$dn/da$}     &  $\chi^{2}/{\rm d.o.f}$ & $\Av$ & $A_{\rm Ks}$  & $\Rv$  & $N_{\rm H}$    \\
          &    silicate    &  graphite       &                         &       &               &        & $\times 10^{22} \rm cm^{-2}$  \\
\hline
\hline \multicolumn{8}{l}{Fitting the observed extinction curve from 1$\um$ to 19$\um$\tablenotemark{a}} \\
\hline
CASE A & \multicolumn{2}{c}{$a^{-2.4}e^{-a/0.04}$}       & 21.3/14 &   39.93 & 2.81  & 2.23 & 7.5\\
CASE B & \multicolumn{2}{c}{$a^{-2.6}e^{-a/0.05}$}       & 20.4/14 &   38.38 & 2.79  & 2.34 & 7.2\\
CASE A & $a^{-3.1}e^{-a/0.10}$ & $a^{-2.7}e^{-a/0.04}$   & 17.2/12 &   40.57 & 2.71  & 2.31 & 7.6\\
CASE B & $a^{-2.9}e^{-a/0.08}$ & $a^{-2.5}e^{-a/0.04}$   & 17.0/12 &   41.28 & 2.69  & 2.34 & 7.7 \\
\hline \multicolumn{8}{l}{Only fitting the observed extinction from 1$\um$ to 7$\um$} \\
\hline
CASE B          & $a^{-3.0}e^{-a/0.09}$ & $a^{-3.1}e^{-a/0.04}$ & 10.5/9 &   39.93 & 2.40 & 2.12 & 7.5 \\
CASE B with AMC & $a^{-3.1}e^{-a/0.12}$ & $a^{-3.0}e^{-a/0.06}$ & 10.4/9 &   35.70 & 2.55 & 2.50 & 6.7 \\
\hline \multicolumn{8}{l}{Only fitting the observed extinction from 3$\um$ to 19$\um$} \\
\hline
CASE B          & $a^{-4.0}e^{-a/0.03}$ & $a^{-3.1}e^{-a/0.02}$ & 10.6/9 &   30.84 & 3.04 & 1.77 & 5.8 \\
CASE B with AMC & $a^{-3.8}e^{-a/0.02}$ & $a^{-3.3}e^{-a/0.03}$ & 11.7/9 &   19.26 & 3.01 & 2.19 & 3.6 \\
\hline \hline
\multicolumn{8}{l}{Fitting with combinations of multi-extinction curves (see $\S$4.3)} \\
\hline
  $f_{\Rv=2.1}$  & $f_{\Rv=3.1}$  & $f_{\Rv=5.5}$ &  $\chi^{2}/{\rm d.o.f}$  & $\Av$ & $A_{\rm Ks}$ & $\Rv$ &   \\
\hline
  0.30 & 0.49 & 0.21                  &  39.1/14 & 34.56 & 2.66 & 2.60 &   \\
  0.28 & 0.39 & 0.33\tablenotemark{b} &  40.7/15 & 33.67 & 2.63 & 2.70 &   \\
\hline

\tablenotetext{a}{We only consider 18 of 21 points of Fritz et al.\ (2011)
  in order to reduce the effect of the 3.1$\mum$ H$_{2}$O feature.}
\tablenotetext{b}{McFadzean et al.\ (1989) argued that the molecular
clouds may contribute as much as $\simali$1/3 ($\sim$10\,mag) of the total visual
extinction $\Av$ towards the GC. Therefore, we fixed the fraction of the $\Rv=5.5$-type extinction to be 0.33.}
\end{tabular}
\label{results}
\end{center}
\end{table}

\section{Model Extinction}\label{fitting}
To testify the dust model, we first fit the standard extinction curve
of $\Rv=3.1$. With $dn/da \propto a^{-3.5}e^{-a/0.14}$ for amorphous
silicates and $dn/da\propto a^{-3.1}e^{-a/0.11}$ for graphite,
the model closely reproduces the $\Rv=3.1$ Galactic extinction curve.
To fit the observed IR extinction curve from 1$\um$ to 19$\um$ toward
the GC (Fritz et al. 2011), for simplicity we first assume that both graphite and
silicate have the same size distribution (i.e. $\alphasil=\alphacarb$,
$\absil=\abcarb$). We then consider models with different power
indices and cutoff sizes for the two dust components to search for
better fits. The best-fit results are summarized in Table~1.
We note that it makes little difference either taking the same size
distribution or assuming different size distributions for silicate and graphite.
None of these attempts could fit the flat MIR extinction well,
although ``CASE B''  works relatively better.

In Figure 2 we show the ``CASE B'' best-fit model extinction
assuming different size distributions for silicate and graphite.
Compared with the observed IR extinction curve toward the GC
(Fritz et al. 2011), the model extinction is too high at the 2.166$\mum$ (Brackett-$\gamma$)
band and too low at $\simali$7$\mum$:
$A_{2.166}^{\rm mod}\approx2.68\magni$
while Fritz et al. (2011) obtained $A_{2.166}\approx2.49\pm0.11\magni$.
The size distribution of $\alphacarb\approx-2.5$
and $\abcarb\approx0.04\um$ for graphite reproduces well
the steep NIR extinction but causes the minimum extinction
near $7\um$. The small cutoff $\abcarb\approx0.04\um$ implies
that the model is rich in small graphite grains so that the model
extinction curve is similar to that of $\Rv=2.1$ in the UV.
The size distribution of $\alphasil\approx-2.9$
and $\absil\approx0.08\um$ for silicate causes
the strong silicate feature at 9.7$\mum$.
Our results show that it may require some dust grains
with a size distribution peaking around 0.5$\um$ or even larger
to produce the flat MIR extinction.
To avoid the complication of the silicate features
we have also modeled the observed extinction but
limiting ourselves to the extinction from 1$\um$ to $7\um$.
To fit the MIR extinction, we have also tried models confining us
to the observed extinction from 3$\um$ to 19$\um$ (i.e., ignoring
the 1--3$\mum$ NIR extinction).
These approaches seem to work well for the chosen wavelength range, but unfortunately,
none of these attempts results in satisfactory fits for the whole range of
1-19$\um$.\footnote{
$^{6}$ Fritz et al. (2011) obtained an optical depth of $\tau_{9.7\mum}\approx 3.84\pm 0.52$ relative
to the continuum at 7$\mum$ from their interpolated extinction curve. However,
in the wavelength range of silicate features, there are too few points to extract the depth
accurately, also because of the large errors. Considering the possible large uncertainty, we
did not use $\tau_{9.7\mum}$ to constraint our fitting.
}
Finally, we replace graphite by amorphous carbon (AMC). But we are
still not able to simultaneously fit both the NIR and MIR extinction.

The NIR extinction law toward the GC derived by Fritz et al. (2011)
and Nishiyama et al.\ (2009) is much steeper than that derived by
Rieke \& Lebofsky (1985) and Rieke et al. (1989), with $\beta \approx -2.0$ compared
to the common value of $\beta \approx -1.6$ to $-1.8$.
For comparison, we also fit the extinction curve of Rieke et al.\ (1989)
, which is actually the $\Rv=3.1$-type extinction, and the model also works very well with
$dn/da \propto a^{-2.1}e^{-a/0.08}$ for amorphous
silicates and $dn/da\propto a^{-3.0}e^{-a/0.28}$ for graphite.
For the sake of clear comparison, we replot in Figure~3
the results shown in Figure~2 but in terms of $A_{\lambda}/\Av$.
We see that the IR extinction toward the GC derived by Fritz et al. (2011)
seems to be a combination of the steep UV-to-NIR extinction of $\Rv=2.1$,
the flat MIR extinction of $\Rv=5.5$,
and the strong silicate feature of $\Rv=3.1$.
It seems that a trimodal size distribution is required in order to
achieve a close fit to the observed extinction from the UV through NIR,
MIR to the silicate absorption band.

\clearpage
\begin{figure}[t]
\centerline{\includegraphics[angle=0,scale=0.5,clip]{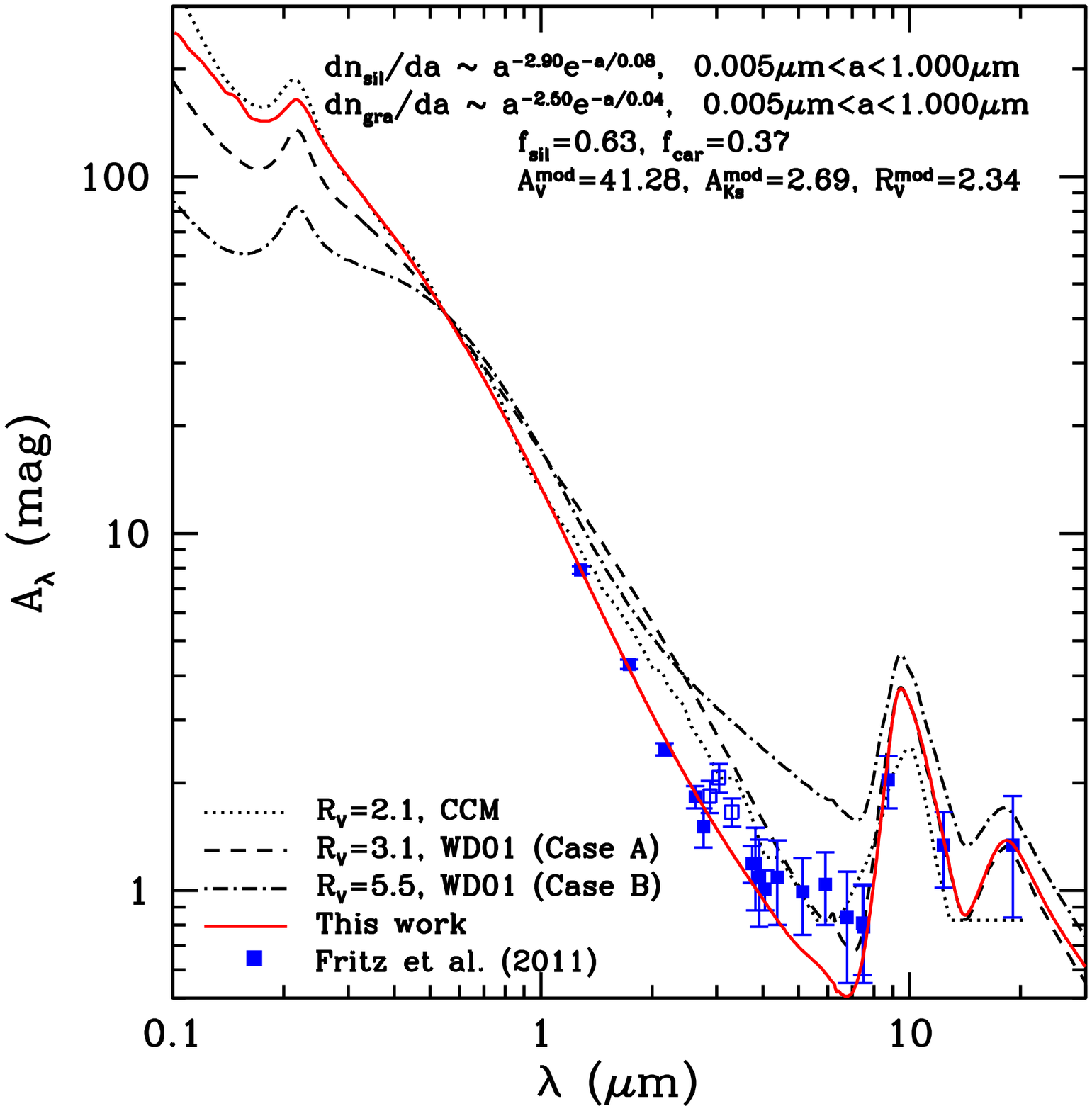}}
\caption{
          Comparison of the model extinction curve (red solid line)
          with the $\simali$1--19$\mum$ IR extinction
          of the GC (blue squares) observed by Fritz et al. (2011).
          Also shown are the extinction curves of
          $\Rv=2.1$ (dotted line, see Cardelli et al.\ (1989)),
          $\Rv=3.1$ (dashed line, WD01)
          and $\Rv=5.5$ (dot-dashed line, WD01)
          with the silicate absorption features added (Draine 2003).
          \label{fig2}
          }
\end{figure}

\clearpage
\begin{figure}[t]
\centerline{\includegraphics[angle=0,scale=0.5,clip]{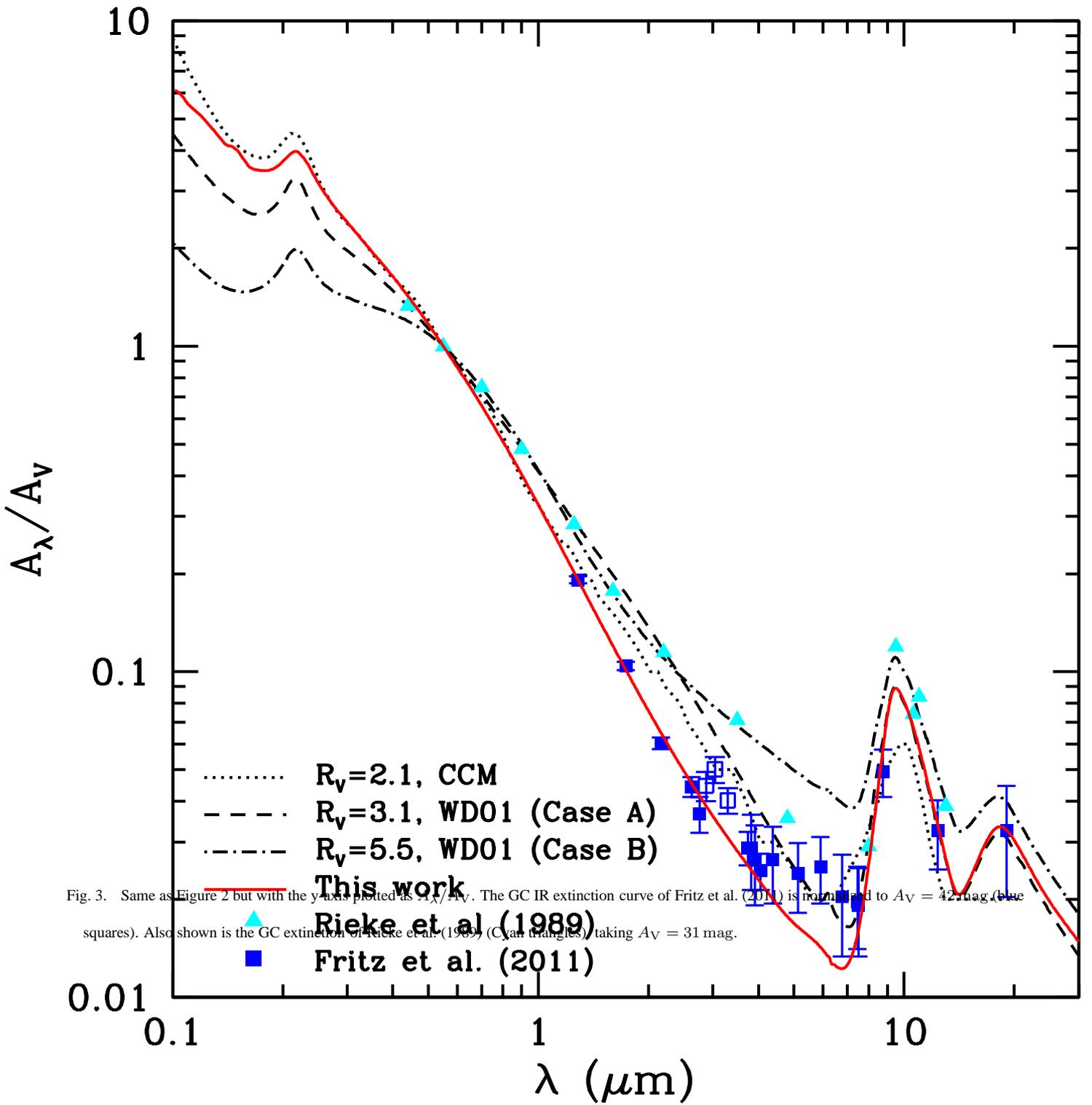}}
\caption{
          Same as Figure~\ref{fig2} but with the y-axis
          plotted as $A_\lambda/\Av$.
          The GC IR extinction curve of Fritz et al.\ (2011)
           is normalized to $\Av=42\magni$ (blue squares).
          Also shown is the GC extinction of Rieke et al.\ (1989) (Cyan triangles),
          taking $\Av=31\magni$.
         \label{fig3}}
\end{figure}
\clearpage

\section{Discussion}\label{discussion}
\subsection{The Extinction Features in the 3--7$\um$ Wavelength Range}
The extinction curve toward the GC obtained by Fritz et al.\ (2011)
shows the strong $3.1\um$ $\rm H_{2}O$ feature
and the 3.4$\um$ aliphatic hydrocarbon feature.
Fritz et al.\ (2011) found that the COMP-AC-S model of Zubko et al.
(2004) seems to best fit their observations as judged by $\chi^2/{\rm d.o.f.}$
and the presence of the H$_2$O ice features. The porosity of ice dust grains
also makes Zubko et al. (2004)'s extinction model fit the GC observed extinction well.
However, the ice features only appear in dense regions, while the flat extinction
in the 3--7$\um$ range is observed towards many
different sightlines, including both diffuse clouds and dense clouds. It is highly possible
that some dust materials other than ices are responsible for the flat MIR extinction towards the GC and elsewhere.\footnote{
$^{7}$ Fritz et al.\ (2011) (see their Section 5.6) argued the flat MIR extinction is not caused
by the molecular clouds in front of the GC, which produce the ice features on the extinction curves. They also
argued (see their Section 5.8) that something else aside from ices produces the flat MIR extinction towards the GC and elsewhere,
and additional pure ice grains produce the extinction features towards the GC.
}
The silicate-graphite dust model considered here is suitable for
the diffuse ISM and does not include ice and aliphatic hydrocarbon
material. Therefore we do not expect to reproduce the 3.1$\mum$
H$_2$O ice feature and the 3.4$\um$ aliphatic C--H feature.

However, these extinction features could be properly reproduced
if the appropriate candidate materials are added in the dust model.
For the 3.4$\um$ aliphatic C--H feature, Draine (2003) argued that
if the graphite component is replaced with a mixture of graphite and
aliphatic hydrocarbons, it seems likely that the extinction curve, including
the 3.4$\um$ feature, could be reproduced with only slight adjustments
to the grain size distribution.
The $3.1\um$ $\rm H_{2}O$ feature may be more complicated
because the $\rm H_{2}O$ feature usually appears in sightlines
passing through dense molecular clouds.
In cold, dense molecular clouds, interstellar dust is expected to
grow through coagulation (as well as accreting an ice mantle)
and the dust is likely to be porous (Jura 1980). Therefore,
introducing a porous structure with ices coated on silicate, graphite
and aliphatic hydrocarbon dust, both the $\rm H_{2}O$ absorption
feature and the 3.4$\um$ aliphatic C--H feature could be reproduced
in the model extinction curves (Zubko 2004; Gao et al.\ 2010).

\subsection{$\Av$: The Extinction at the Visual Band}
Rieke et al.\ (1989) estimated the visual extinction toward the GC
to be $\Av\approx 31\magni$ based on the extinction law of
Rieke \& Lebofsky (1985) ($\Rv=3.1$). Our best-fit model for the
Rieke et al.\ (1989) extinction law also gives
$\Av \approx 31.4\magni$.
However, with $\beta \approx-2.11 \pm 0.06$,
Fritz et al. (2011) obtained $\Rv\approx2.48 \pm 0.06$
for the extinction toward the GC based on the correlation
between $\Rv$ and the IR power-law index $\beta$ of Fitzpatrick \& Massa (2009).
Fritz et al. (2011) obtained $\Av \approx 44\magni$ by extrapolating this
curve. They also argued that the X-rays can shed lights on $\Av$,
and $\Av$ toward the GC may be higher,
up to $\simali$59$\magni$ (assuming different $N_{\rm H}/\Av$ ratios).

Our model extinction curves suggest that models for small $\Rv$ ratios
work better for the steep NIR extinction obtained by Fritz et al. (2011).
Since a smaller $\Rv$ ratio implies a higher $\Av$ (on a per unit
NIR extinction basis), this again suggests that $\Av$ toward the GC
is probably larger than previous estimated.
Our best-fit models suggst that $\Av$ toward the GC is
$\simali$42$\magni$ (see Table~1).
If we do not fix the total silicate ($\Vsil$) and graphite volume
($\Vcarb$), instead, we allow the quantity of the silicate
component to vary with respect to that of graphite:
by taking the silicate-to-graphite mass ratio to be
$m_{\rm gra}/m_{\rm sil}=$ 0.4, 0.5, and 0.6, our model results
show that $\Av$ toward the GC is in the range of
$\simali$35--45$\magni$.
In the diffuse ISM, $\Av/N_{\rm H} \approx5.3 \times10^{-22}\magni\cm^{2}$ (WD01), which
leads to $N_{\rm H}\approx 7.7 \times10^{22}\cm^{-2}$ for our best
``CASE B'' model extinction curve. However, towards the GC,
the interstellar environments should be much denser than that of the diffuse ISM.
Although $\Av/N_{\rm H}$ is less clear for dense clouds,
Cardelli et al.\ (1989) and Draine (1989) argued that
$A_{\rm I}/N_{\rm H}\approx2.6 \times 10^{-22}\magni\cm^{2}$
typical of the diffuse ISM may also hold for dense clouds.
If this is indeed the case, we estimate the
column density $N_{\rm H}$ for the sightline toward the GC
to be $N_{\rm H} \approx 6.42 \times10^{22}\cm^{-2}$ for our best
``CASE B'' model extinction curve ($A_{\rm I}\approx16.69\magni$).
It is smaller than $N_{\rm H}\approx\left(10.5\pm1.4\right)\times10^{22}\cm^{-2}$
obtained by Fritz et al. (2011) which implies $\Av/N_{\rm H}\approx6.6 \times 10^{-22}\magni\cm^{2}$.
It is also much smaller than that of Nowak et al. (2012), who derived the X-ray absorbing
column density to be $N_{\rm H}\approx15\times10^{22}\cm^{-2}$.

\subsection{A Simple Model Based on
            Combinations of Multi-Extinction Curves}
When the starlight from the GC reaches us,
it may have passed through the spiral arms
where star formation is actively occurring,
diffuse regions, and dense regions of molecular clouds.
McFadzean et al.\ (1989) argued that the molecular clouds
along the line of sight toward the GC may contribute
as much as $\simali$1/3 ($\sim$10\,mag)
of the total visual extinction $\Av$.
Therefore, the extinction curve toward the GC
may be a combination of different extinction curves
produced by dust grains in different environments
of different size distributions.
The best fits of this trimodal model are shown in the last two rows of Table 1.
The first row shows the best fit derived by varying the contribution of
different extinction curves (i.e. $\Rv$), while the 2nd row is for fixing the $\Rv=5.5$-type
extinction to account for 1/3 of the total extinction if we assume the molecular cloud contributes
as much as $\simali$1/3 ($\sim$10\,mag) of $\Av$ towards the GC.
As shown in Figure~4, the observed IR extinction of the GC
is fitted well in terms of three different extinction curves,
characterized by $\Rv=2.1$, 3.1, and 5.5, respectively, each contributing
30\%, 49\%, and 21\% of the total $\Av$, with the $\Rv=2.1$ extinction representing that of the region where the
dust subjects to heavy processing such as in HD\,210121, a high Galactic
latitude cloud (Larson et al.\ 1996, Li \& Greenberg 1998).\footnote{
$^{8}$ In Figure 4, although it appears to fit the extinction well in the range of 1.2-8.0$\mum$,
the $\Rv=2.1$ (HD\,210121) extinction curve actually is not the suitable extinction curve for the interstellar
environment towards the GC because it predicts a very strong silicate absorption feature at 9.7$\mum$.}
Although the $\chi^2$ is not lower than that of single $\Rv$ models (see Table 1), we think that
the trimodal model is an useful description because it seems reasonable that the dust in the lines of sight
towards the GC is characteristics of different environments.

%


\begin{figure}[t]
\centerline{\includegraphics[angle=0,scale=0.5,clip]{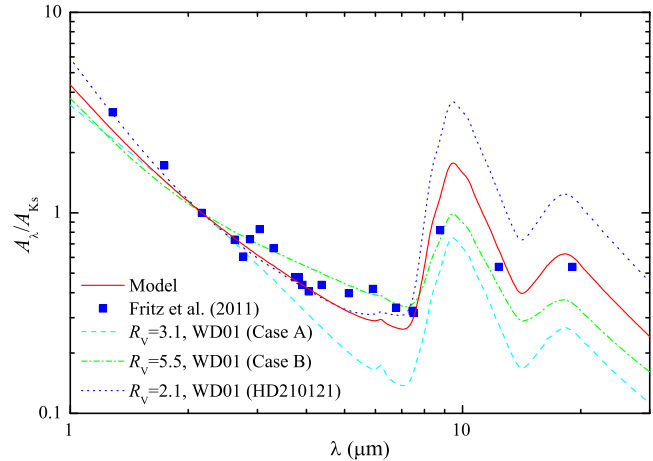}}
\caption{
             Comparison of the GC IR extinction of
             Fritz et al.\ (2011; blue squares)
             with the best fit model extinction (red solid line)
             obtained from the combination of three extinction curves
             of $\Rv=2.1$ (28\%, dotted line),
             $\Rv=3.1$ (39\%, dashed line),
             and $\Rv=5.5$ (33\%, dot-dashed line). Also shown are the observed extinction toward the GC by Nishiyama et al.\ (2009; green triangles)
             and the best fit with our dust model (see $\S$3; red dotted line).
             \label{fig4}}
\end{figure}

\section{Summary}\label{summary}
The $\simali$1--19$\mum$ IR extinction curve of the GC
recently derived by Fritz et al.\ (2011) is fitted with
a mixture of graphite and amorphous silicate dust.
%
The model has difficulty in simultaneously reproducing
the steep NIR extinction and the flat MIR extinction.
The best-fit model estimates the total visual extinction
toward the GC to be $\Av \sim 38 - 42 \magni$.
%
In view that the starlight from the GC passes through different
interstellar environments, the observed extinction curve toward
the GC could be a combination of different extinction curves
produced by grains with different size distributions characteristic of different environments:
dust in diffuse regions (characterized by small $\Rv$ and steep near-IR extinction),
and dust in dense regions (characterized by large $\Rv$ and flat UV extinction).¡±


\acknowledgments{We thank the anonymous referees for their comments that
helped improve the presentation of the paper. This work is supported by NSFC grant No.\,11173007,
NSF AST 1109039, and the University of Missouri Research Board.
This publication was also made possible through the support of
a grant from the John Templeton Foundation. The opinions expressed
in this publication are those of the authors and do not necessarily
reflect the views of the John Templeton Foundation. The funds from
John Templeton Foundation were awarded in a grant to
The University of Chicago which also managed the program
in conjunction with National Astronomical Observatories,
Chinese Academy of Sciences.}



\email{J.~GAO (e-mail: jiangao@bnu.edu.cn), A. Li (e-mail: lia@missouri.edu), and B.~W.~Jiang (e-mail: bjiang@bnu.edu.cn)}
\label{finalpage}
\lastpagesettings
\end{document}